# Stephan Prantner's Sunspot Observations during the Dalton Minimum

Hisashi Hayakawa (1–4), Shoma Uneme (1), Bruno P. Besser (5–6), Tomoya Iju (7), Shinsuke Imada (1)

(1) Institute for Space-Earth Environmental Research, Nagoya University, 4648601, Nagoya, Japan

(2) Institute for Advanced Researches, Nagoya University, 4648601, Nagoya, Japan

(3) UK Solar System Data Centre, Space Physics and Operations Division, RAL Space, Science and Technology Facilities Council, Rutherford Appleton Laboratory, Harwell Oxford, OX11 0QX, Didcot, Oxfordshire, UK

(4) Nishina Centre, Riken, 3510198, Wako, Japan

(5) Space Research Institute, Austrian Academy of Sciences, 8042, Graz, Austria

(6) Institute of Physics, University of Graz, 8010 Graz, Austria

(7) National Astronomical Observatory of Japan, 1818588, Mitaka, Japan

**Abstract**

In addition to regular Schwabe cycles (≈ 11 years), solar activity also shows longer periods of enhanced or reduced activity. Of these, reconstructions of the Dalton Minimum provide controversial sunspot group numbers and limited sunspot positions, partially due to limited source record accessibility. We analysed Stephan Prantner's sunspot observations from 1804–1844, the values of which had only been known through estimates despite their notable chronological coverage during the Dalton Minimum. We identified his original manuscript in Stiftsarchiv Wilten, near Innsbruck, Austria. We reviewed his biography (1782–1873) and located his observational sites at Wilten and Waidring, which housed the principal telescopes for his early and late observations: a 3.5-inch astronomical telescope and a Reichenbach 4-feet achromatic erecting telescope, respectively. We identified 215 days of datable sunspot observations, which are twice as much data as his estimated data in the existing database (= 115 days). Prantner counted up to 7–9 sunspot groups per day and measured sunspot positions, which show their distributions in both solar hemispheres. These results strikingly emphasise the difference between the Dalton Minimum and the Maunder Minimum as well as the similarity between the Dalton Minimum and the modern solar cycles.

**1. Introduction**

The numbers of sunspot groups and individual sunspots — which have been monitored since 1610 — form a direct basis for evaluating the magnitude of solar activity for ≈ 410 years, as visual representations of solar magnetic fields (Vaquero and Vázquez, 2009; Clette *et al*., 2014; Arlt and Vaquero, 2020). These observations form one of the longest ongoing scientific experiments in human history (Owens, 2013). These data have identified several cyclicities, such as the regular ≈ 11-year Schwabe cycles (Hathaway, 2015). These data show fairly good correlations with other physical measurements such as sunspot area and solar-irradiance fluxes, and hence used for their cross-comparisons, reconstructions, and recalibrations (Hathaway, 2015; Tapping and Morgan, 2017; Criscuoli *et al*., 2018; Chatzistergos *et al*., 2019, 2020; Clette, 2020; Lean *et al*., 2020). This is also the case with the proxy data like cosmogenic isotopes, such as $^{14}$C in tree rings, $^{10}$Be in ice cores, and $^{44}$Ti in meteorites (Solanki *et al*., 2004; Beer *et al*., 2012; Usoskin, 2017; Asvestari *et al*., 2017). These correlations have often used to chronologically extend the solar-irradiance fluxes beyond their observational onsets (*e.g.*, Ermolli *et al*., 2013; Lean, 2017; Berrilli *et al*., 2020).

Two peculiar 'prolonged solar minima' — the Maunder Minimum (1645–1715) and the Dalton Minimum (1797–1827) — have been identified within the multi-century sunspot observations (*e.g.*, Usoskin, 2017). The Maunder Minimum was named after Edward W. Maunder (1922), who noticed significantly reduced sunspot occurrences, reduced auroral frequency, and increased cosmogenic isotopes (Eddy, 1976; Usoskin *et al*., 2015). This peculiar period has attracted significant scientific interest and has been characterised by suppressed solar cycles (Vaquero *et al*., 2015; Usoskin *et al*., 2015), asymmetric sunspots (Ribes and Nesme-Ribes, 1993; Muñoz-Jaramillo and Vaquero, 2019), and apparent loss of solar coronal streamers (Riley *et al*., 2015; Owens *et al*., 2017; Hayakawa *et al*., 2021). This period is generally considered a grand solar minimum and used as a standard reference for other grand minima identified in the proxy data of cosmogenic isotopes (Solanki *et al*., 2004; Usoskin *et al*., 2007, 2014; Inceoglu *et al*., 2015; Muscheler *et al*., 2016; Brehm *et al*., 2021).

The other peculiar 'prolonged solar minimum' in the direct sunspot observations has been named as the Dalton Minimum after John Dalton (Dalton, 1834; Silverman and Hayakawa, 2021). Its relationship with the Maunder Minimum has been somewhat controversial. Some have expected similar physical mechanism (*e.g.*, Kataoka *et al*., 2012; Zolotova and Ponyavin, 2015), whereas others have considered them significantly different in physical mechanism (Vaquero *et al*., 2015;

Usoskin *et al*., 2015; Charbonneau, 2020; Petrovay, 2020). Challengingly, as reviewed in Figure 2 of Muñoz-Jaramillo and Vaquero (2019), the composite data series of currently available sunspot group numbers significantly varied around the Dalton Minimum (*e.g.*, Lockwood *et al*., 2014; Clette and Lefevre, 2016; Svalgaard and Schatten, 2016; Usoskin *et al*., 2016; Chatzistergos *et al*., 2017), even after intensive revisions and recalibrations of sunspot group numbers (Clette *et al*., 2014; Vaquero *et al*., 2016) following the initial compilations in Hoyt and Schatten (1998a, 1998b = HS98). Even worse, little has been known about sunspot positions during the Dalton Minimum, whereas its peculiarity has significant indications to the long-term solar variability and its background solar-dynamo activity (*e.g.*, Usoskin, 2017; Muñoz-Jaramillo and Vaquero, 2019; Charbonneau, 2020). As all the existing solar-irradiance reconstructions during the Maunder Minimum and the Dalton Minimum have relied on sunspot data or cosmogenic isotopes, their revisions and recalibrations potentially improve the reconstructions for solar-irradiance fluxes and their impact estimates for the terrestrial climate (*e.g.*, Solanki *et al*., 2013; Anet *et al*., 2014; Kopp *et al*., 2016; Lean, 2017; Berrilli *et al*., 2020).

The initial difficulties are partly due to the limited accessibility to the original — and often unique –– source documents. Considerable amounts of long-term observational records are preserved in the historical archives and hence difficult to access, as documented in the bibliographic supplements in HS98. However, understanding of sunspot activity during the Dalton Minimum has gradually improved with new observational evidence and efforts to recalibrate the sunspot group number (Clette *et al*., 2014). Short-term observations have been added to the existing datasets (Denig and McVaug, 2017; Hayakawa *et al*., 2018; Carrasco *et al*., 2018), and the original manuscripts of Derfflinger's long-term observations from 1802–1824 have been reanalysed to revise sunspot group numbers and derive sunspot positions to emphasise contrasts with the Maunder Minimum (Hayakawa *et al*., 2020a). Significant solar coronal streamers were recorded during the Dalton Minimum (Hayakawa *et al*., 2020b), unlike the apparent loss of the solar coronal streamers during the Maunder Minimum (Eddy, 1976; Riley *et al*., 2015; Hayakawa *et al*., 2021). Still, these analyses still further evidence, especially because only Derfflinger's data series during the Dalton Minimum has been fully studied both in terms of the long-term sunspot group number and sunspot positions.

In this context, Stephan Prantner's intermittent but long observations (1804–1844) provide a series of interesting contextual datasets during the Dalton Minimum and relatively rich observational data (115 days). However, Prantner's data in the existing datasets (HS98; V+16) are not constructed from the original documents but estimated from his individual spot number ($f$) and other observers' data at that time. HS98 themselves explicitly admitted that they could not access the entire data in their online bibliography: "We have seen some of these observations, but may not have all of them in our database". This is further confirmed by Douglas Hoyt's letter correspondence with Fritz Steinegger of the Stiftsarchiv Wilten on 13 May 1994, which is registered as MS 07 03 20 in this archive. Here, Douglas Hoyt confirmed to have received "the number of individual sunspots and not the number of sunspot groups" and "constructed estimates of the number of sunspot groups based on other observers", whereas "copies of the sunspot drawings could not be made and sent to me [= Douglas Hoyt]". This letter correspondence indicates that Prantner's data in the existing dataset is not empirical data acquired from the original manuscript but estimates with his individual sunspot number ($f$) and contemporary observations. These estimates were acquired in the revised database (V+16) without revision. Therefore, we consulted Prantner's original manuscripts in the Stiftsarchiv Wilten and derived the sunspot group number and sunspot positions on their basis. Section 2 has reviewed Prantner's personal profile and observational method. Section 3 has derived his daily sunspot group number and compared them with contemporary observations both in daily and annual bases. Section 4 has derived sunspot positions from his manuscripts and constructed butterfly diagram. Section 5 has summarised these results and compared them with the known data in the Maunder Minimum and the modern solar cycles.

## 2. Prantner's Personal Profile and Observational Methods

Stephan Prantner was a clergyman and teacher who worked in Tyrol. He was born on 23 June 1782 in Innsbruck and died on 18 May 1873 in Wilten. He studied at Innsbruck, particularly preferred mathematics and natural science, and joined a student legion against the French invasion in 1799. He later joined the Premonstratensians at Wilten Monastery (Figure 1) in 1800, took the vow in 1803, and was consecrated as a priest at Brixen in 1805. After its abolition by the Bavarian government in 1807, he settled as a vicar in the Tyrolian municipality Waidring near Kitzbühel in the archdiocese of Salzburg after having served as a priest in several other Tyrolian villages during his Wilten time. He returned to Wilten in 1816. He was appointed professor of physics and mathematics and later from 1836 subprior of the Wilten Monastery. Throughout his career, Prantner worked on astronomy, mineralogy, and botany, and conducted meteorological observations for almost half a century (Anonymous, 1873; Kosch, 1937; Stift Wilten, 1989).

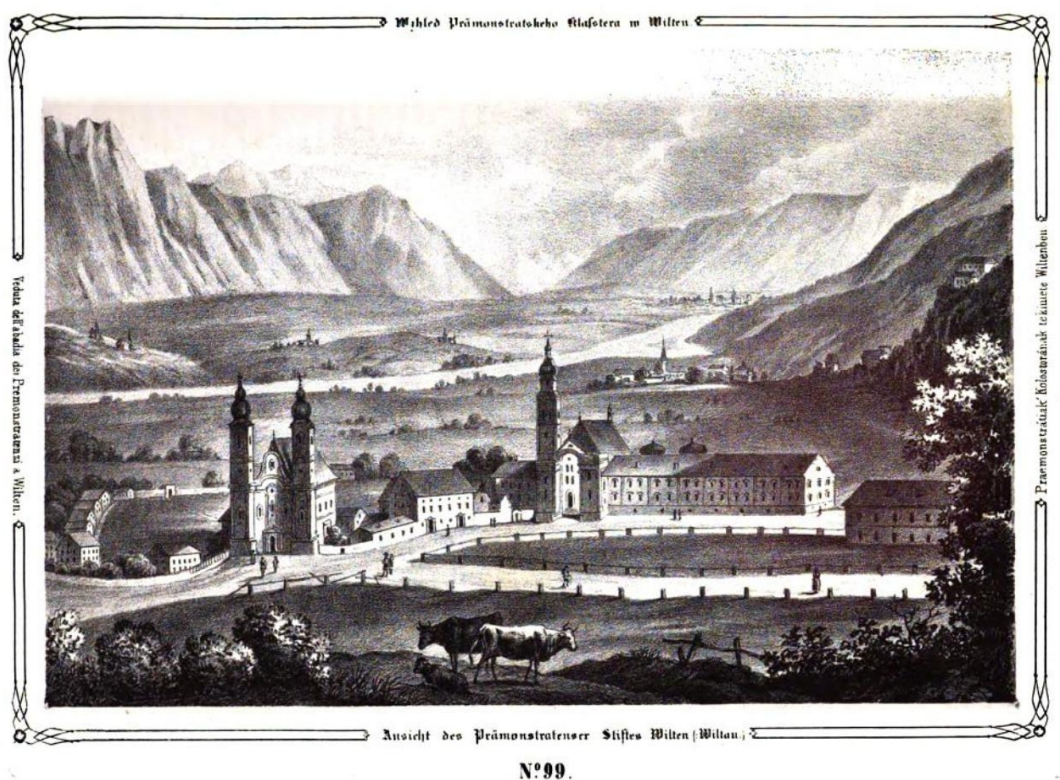

Figure 1: Wilten Monastery in Prantner's epoch depicted in Plate No. 99 of Ziegler (1843).

His manuscripts are currently preserved in Stiftsarchiv Wilten (Figure 1). Of these, Prantner's sunspot drawings are preserved as a small pamphlet in MS A07 03 07. This manuscript is entitled *Zu den Hefften der Tagebücher | Physisch-Astronomischer Beobachtungen und Ereignisse. | Beobachtete Mittags Sonnen-Höhen im Stifte Wilten | von Stephan Prantner. Canonie. Wilten*, which reads: "The journal diary of | physical-astronomical observations and events. | Observations of midday solar altitudes in Stift Wilten | by Stephan Prantner cannonicus at Wilten". His pamphlet has two sections, as exemplified in Figure 2. The first section is entitled *Abbildungen | beobachteter Sonnen-Flecken | vom Anfang des Jahrs 1812 bis Ende | des Jahrs 1815 | beobachtet und gezeichnet von Stephan Moritz Prantner*, which reads "Drawings of observed sunspots from the beginning of 1812 to the end of 1815 observed and drawn by Stephan Moritz Prantner". The second section is entitled *Beobachtete und gezeichnete Abbildungen der Sonnenflecken von Stephan Prantner Canonic. Wiltin*, which reads "Observed and drafted images of sunspots by Stephan Prantner canonicus Wilten"

Figure 2: Examples of Prantner's sunspot drawings; (a) the figure above shows his early sunspot drawings from 23–24 September 1813 (MS A07 03 07, *f.* 6); (b) the figure below shows his late

drawings on 9 April 1816 and 5 October 1816 (MS A07 03 07, *f.* 16 and *f.* 27b). These images are reproduced with courtesy of the Stiftsarchiv Wilten.

The sunspot observations in these pamphlets were dated to: 1804 and 1812–1815 (the first pamphlet); and 1816, 1826–1827, and 1844 (the second pamphlet). Prantner clarified his observational sites as Wilten (N47°15, E11°24) in 1804 (MS A07 03 07, f. 1), Waidring (N47°35, E12°34) in 1812–1815 (MS A07 03 07, ff. 2–13), and Wilten after 1816 (MS A07 03 07, f. 14). The described transition of his observational sites is broadly consistent with his career reviewed above (Anonymous, 1873). Prantner explained the observational gap from 1804–1812 as due to time and circumstantial difficulties as well as his residential movements in this interval (MS A07 03 07, f. 1).

Prantner used at least two telescopes. His early observations (1804–1815) were performed with a 3.5-inch astronomical telescope; the sunspots observed through the telescope were drafted with inverted images (MS A07 03 07, f. 0b). Prantner used a Reichenbach 4-feet achromatic erecting telescope for his latter observations and depicted sunspots in erect images, unless otherwise endorsed (MS A07 03 07, f. 15). Further details are difficult to locate, as some of his early instruments and books seemed to have been lost during the French invasions (Anonymous, 1873).

## 3. Sunspot Group Number

We consulted Prantner's manuscript (MS A07 03 07) and identified his sunspot observations for 215 days in 1804, 1812–1816, 1826–1827, and 1844. We have summarised these data in: https://www.kwasan.kyoto-u.ac.jp/~hayakawa/data/. In comparison with the existing data (= 115 days) in HS98 and V+16, our data have almost doubled Prantner data. This drastic difference has been caused because Prantner's manuscripts had not been directly consulted in previous studies but estimated from his individual sunspot number (*f*) and other contemporary data, as confirmed in Douglas Hoyt's correspondence to Steinegger of the Stiftsarchiv Wilten in 1994 (MS 07 03 20). Our analyses thus form brand new data from Prantner's original manuscripts.

Prantner's observations were most active in Solar Cycle 6 (August 1810 to May 1823; Table 2 of Hathaway, 2015) and covered 5 days in 1804, 41 days in 1812, 23 days in 1813, 15 days in 1814, 41 days in 1815, 83 days in 1816, 4 days in 1826, 1 day in 1827, and 2 days in 1844. Our analyses will increase the temporal coverage of the data in 1812 (+3 days), 1815 (+16 days), and 1816 (+80 days) and will slightly improve the temporal coverage of V+16.

We applied the Waldmeier classification (Kiepenheuer, 1953) to derive the sunspot group number from the original manuscripts and summarised our results in Figure 3. Here, we have derived both of his sunspot group number and individual sunspot number but concentrated our discussions on his sunspot group number, as his descriptions for individual spots are significantly different in his early observations and late observations. Prantner's observations (Figure 3) are classified into two categories: 1804–1815 (Figure 2a) and 1816–1844 (Figure 2b). Their depictions are significantly different from each other, partially because of the use of different telescopes. Prantner recorded up to 7 groups in the former half (6 July 1815) and 9 groups in the latter half (5–6 October 1816). Prantner's record on 6 July 1815 is consistent with von Lindener's observations on 5–6 July 1815 (7 groups in V+16), whereas his data on 5–6 October 1816 significantly exceed von Lindner's on 5 October 1816 (5 groups in V+16). Prantner recorded only one spotless day (24 August 1816). Additionally, these data explicitly show the ascending phase of Solar Cycle 6 from 1812–1816. Thus, it is notable to identify a spotless day in 1816, immediately after the maximum in May 1816. This spotless day is also confirmed in Tevel's data on the same date (V+16).

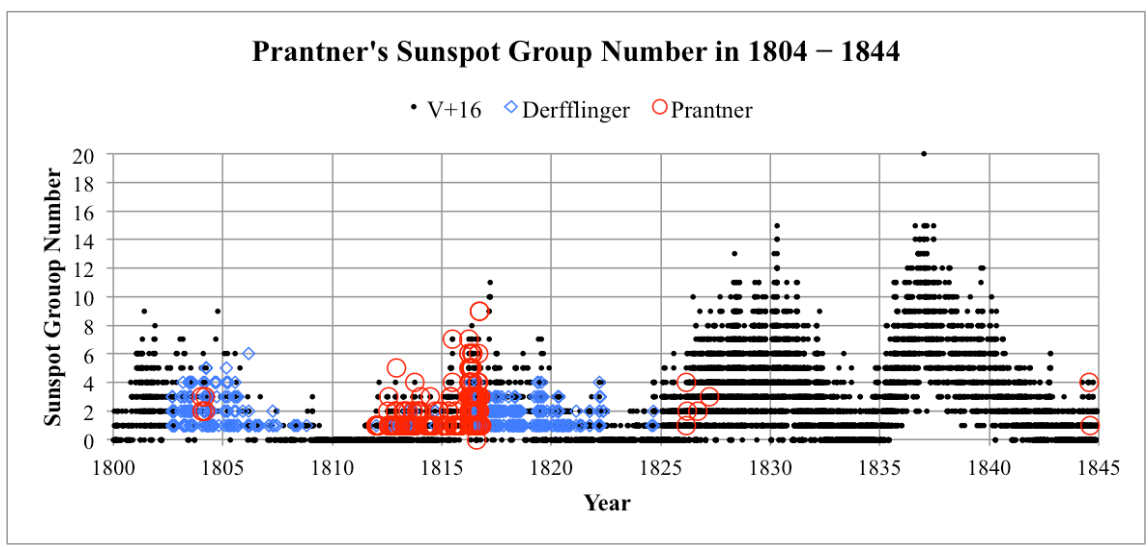

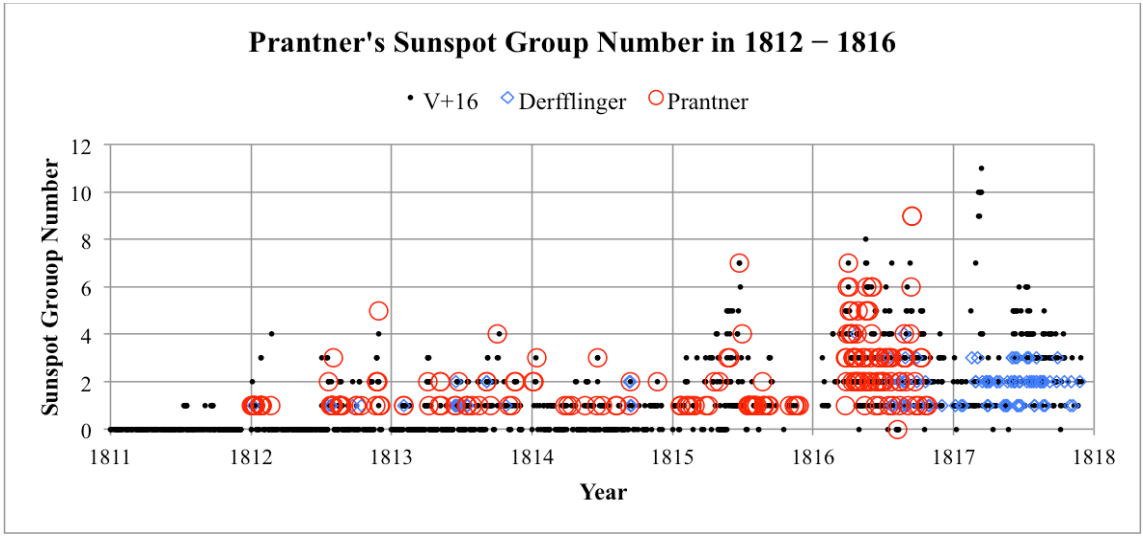

Figure 3: Prantner's sunspot group number (in red) compared with contemporary observations: the existing observers (V+16) in black and Derfflinger (Hayakawa *et al*., 2020a) in blue; (a) The figure above shows this comparison for Prantner's entire interval (1804–1844) with the other observations in 1800–1844; (b) the figure below shows Prantner's core period (1812–1816) with other observations in 1810–1819.

We have computed the annual averages of Prantner's sunspot group number and compared them with those of Derfflinger's sunspot group number (Hayakawa *et al*., 2020a) and the yearly sunspot group number in Chatzistergos *et al*. (2017). For the group sunspot number data, we have derived the error margins with standard deviations of individual data, as done in the SILSO[1]. Their results are summarised in Figure 4. On their basis, Prantner's sunspot group number is almost comparable with or slightly larger than Derfflinger's in his early observations (1804 and 1812–1815), whereas Prantner's late observations are significantly larger than Derfflinger's in 1816.

We have also compared their common observations on the same dates. Here, we have detected 24 days of their overlapping observations on the same dates. On this basis, the annual averages of Prantner and Derfflinger turn out: 2.5 ± 0.5 vs 1.5 ± 0.5 in 1804, 1 vs 1 in 1812, 1.5 ± 0.2 vs 1.5 ± 0.2 in 1813, 1.5 ± 0.5 vs 2 in 1814, and 2.5 ± 0.2 vs 2.3 ± 0.3 in 1816. This result implies the Prantner's sunspot group number agree with Derfflinger's simultaneous observations within standard deviations. In this case, enhancements in Prantner's sunspot group number in 1816 may be considered real rather than instrumental issues, as his observations were much richer in 1816 (for 83 days). In fact, Prantner reported the Sun with a significantly increased sunspot activity on the days in April – June and early October 1816, when Derfflinger did not conduct observations (see Hayakawa *et al*., 2020a). These intervals include 5–6 October 1816, when Prantner reported 9 sunspot groups, for example. Considering the different reconstructions of sunspot group number series over that period (*e.g.*, Figure 5 of Chatzistergos *et al*. (2017)), these results seem more consistent with those of Usoskin *et al*. (2016) and Chatzistergos *et al*. (2017), in contrast to HS98 and Svalgard and Schatten (2016). Therefore, it is possible that SC6 was slightly larger than SC5, as recorded in Prantner's observations. This requires us to examine further long-term observers' data during the Dalton Minimum in terms of their data and observational methods.

---

[1] http://sidc.oma.be/silso/infosnytot

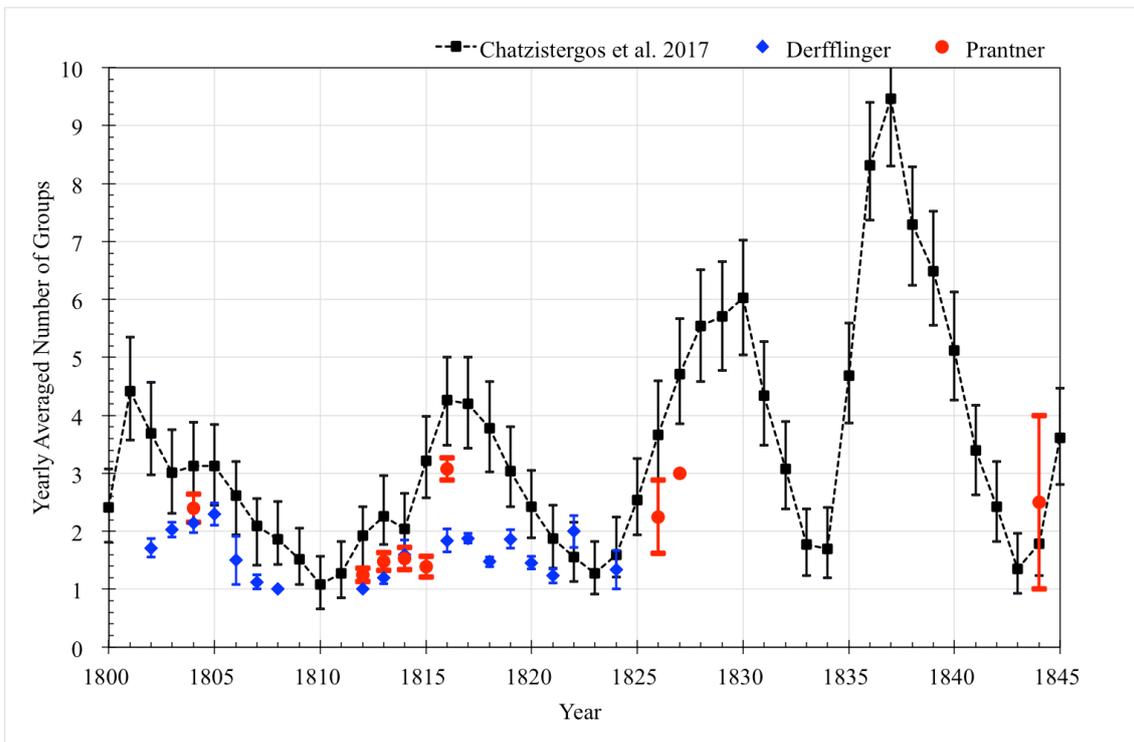

Figure 4: Annual averages of Prantner's sunspot group number in comparison with those of Derfflinger's sunspot group number (Hayakawa *et al*., 2020a) and the yearly sunspot group number (Chatzistergos *et al*., 2017).

Caveats must be noted here. Except for his core observations in 1812 – 1816, his observations (those in 1804, 1826 – 1827, and 1844) involve only ≤ 5 days for each year and cannot fully represent their annual averages. As also shown here, both of Prantner and Derfflinger show their annual sunspot group number ≥ 1. This is because spotless days were reported only once in Prantner's observations and not in Derfflinger's. Therefore, Prantner's report for a spotless day in 1816 is extremely important and their annual averages have been slightly exaggerated due to the omissions of the spotless days. These caveats require us to be cautious for the exaggeration of the solar-cycle amplitude around the Dalton Minimum, whereas their daily data still form ground truth in contrast with their annual averages.

**4. Sunspot Positions**

We measured sunspot positions in Prantner's sunspot drawings and constructed a butterfly diagram. The orientations of the sunspot drawings in the early (1804–1815) and late (from 1816–1844) observations vary. Prantner's manuscript explains his telescopes showing an inverted image in the early observations and erect image in the late observations (MS A07 03 07, *f*. 0b and *f*. 15). This interpretation is confirmed with the sunspot motions from right to left and from left to right in the early and late drawings, respectively. Therefore, the early drawings must be inverted both vertically and horizontally, and the late drawings should be compared with the solar disk as seen in the sky in terms of their orientations. Prantner also described the raw disk orientations as viewed through his telescopes in the folio margins of his manuscript for his early drawings and at the first sunspot drawing for his late observations.

We inferred Prantner's disk orientations with the recorded local time and sequential motion of individual sunspot groups. Prantner recorded time stamps for each sunspot drawing with local time (MS A07 03 07, f. 0b) and recorded the sequential motion of the same sunspot groups, probably due to his interest in solar rotation. We used two methods to fix the position angle of the solar disks in Prantner's manuscript. For sunspot drawings recorded in a chronological sequence, we tracked the same sunspot groups and minimised their latitudinal variations with rotational matching (see *e.g.*, Arlt *et al*., 2013). For chronologically isolated sunspot drawings, we assumed the position angle from the described local time using the ephemeris of JPL DE430 (Folkner *et al*., 2014).

Fixing the position angle of Prantner's sunspot drawings, we measured the heliographic coordinates of the depicted sunspot groups. Note that Prantner also made position measurements for some of the observed sunspots and solar radius on some dates, whereas they do not cover all his sunspot observations (Figure 2). Especially, in his late observations, Prantner's positions measurements are limited for specific groups and not for solar diameter. Therefore, as a matter of consistency, we measured the sunspot positions from Prantner's drawings for the entire period. Here, we have used our own photographs, in order not to damage the original manuscripts (see also MS 07 03 20). In order to minimise the geometrical distortion, we have manually detected the described disk limbs and geometrically fitted the depicted disk limbs to circles following the procedures in Fujiyama *et al*. (2019). We have derived the apparent sunspot positions in the sunspot drawings with circle fitting and modified their disk orientations following the position angle for each drawing.

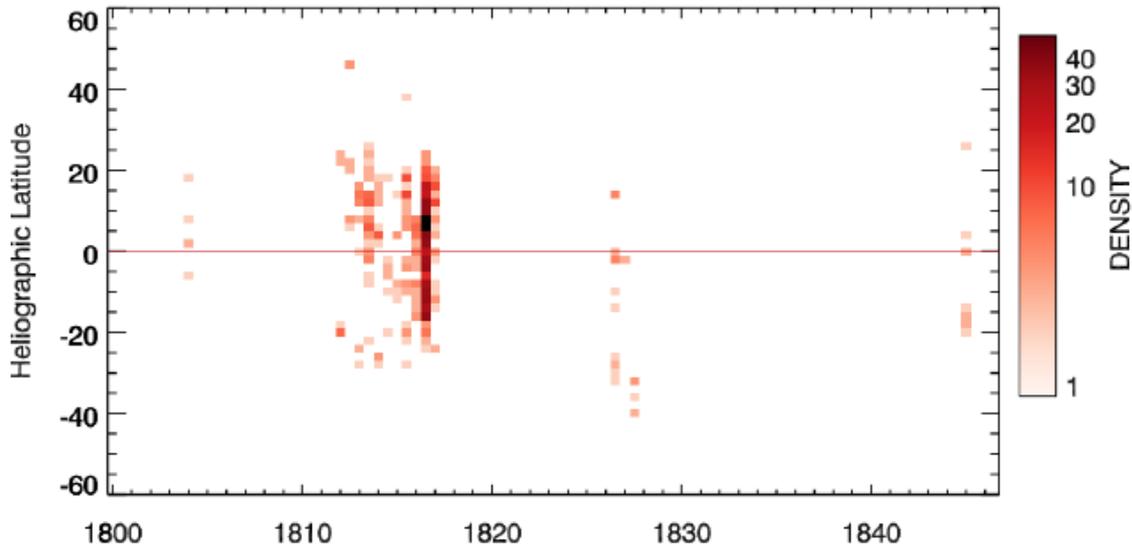

Figure 5: Butterfly diagram constructed from Prantner's manuscripts. Each bin has been set for a 6-month period in chronological order and set to 1° of heliographic latitude. The colourations show data density.

We measured the heliographic coordinates of each sunspot group, constructed butterfly diagrams (Figure 5), and showed the sunspot population of each bin for 6-month periods in chronological order for 1° heliographic latitude increments, and marked their concentrations in blue to red. The populations of sunspot groups in significantly high heliographic latitudes (≈ |50|°) are significantly small within the entire dataset and could be attributed to lower accuracy. However, the majority of sunspots remain within |30|° and provide insights on Solar Cycles 5, 6, 7, and 9.

Prantner's butterfly diagram immediately visualises sunspot distributions in both solar hemispheres. These results are consistent with Derfflinger's sunspot drawings during 1802–1824 (Hayakawa *et al*., 2020a) and contrasted with those of the Maunder Minimum, when sunspot groups were mostly concentrated in the southern solar hemisphere (Ribes and Nesme-Ribes, 1993; Muñoz-Jaramillo and Vaquero, 2019). Prantner's core observations from 1812–1816 show slight equatorial migrations of recorded sunspots and agree with the chronological context of the ascending phase from August 1810 and the maximum in May 1816 (Tables 1–2 of Hathaway, 2015). Prantner's observations in 1804, 1826–1827, and 1844 are situated immediately before the maximum of Solar Cycle 5 in February 1805, in the ascending phase of Solar Cycle 7 in November 1829, and immediately after the minimum of Solar Cycle 8/9 in July 1843, respectively (Tables 1–2 of Hathaway, 2015). These chronological contexts are consistent with the relatively large latitudinal scattering of the observed sunspot groups in these years.

## 5. Summary and Discussions

In this article, we analysed Stephan Prantner's sunspot observations. We first reviewed his life from 1782–1873. Prantner worked in the Premonstratensians at Wilten Monastery beginning in 1800, received formal education in mathematics and astronomy, and became a professor of physics and mathematics and subprior of Wilten Monastery in 1836.

Prantner's sunspot observations are currently preserved in the MS A07 03 07 of the Stiftsarchiv Wilten and are dated to 1804, 1811–1816, 1826–1827, and 1844. We located his observational sites as Wilten (N47°15, E11°24) in 1804, Waidring (N47°35, E12°34) in 1812–1815, and Wilten from 1816 onward, and clarified that Prantner primarily used a 3.5-inch astronomical telescope in his early observations (1804–1815) and a Reichenbach 4-feet achromatic erecting telescope in his later observations to show erect and inverted solar images, respectively.

We consulted his manuscripts, identified 215 days of datable sunspot observations, and doubled the data in comparison with his group number in HS98 and V+16 (= 115 days), which were estimated from Prantner's individual spot number (*f*) and contemporary observers' group number in HS98. We applied the Waldmeier classification and derived the sunspot group number (Figure 3). Prantner recorded up to 7 and 9 groups in his early and late observations, respectively, and reported only one spotless day in 1816, immediately after the maximum of SC6. We also confirmed that Prantner's observations are generally consistent with contemporary observations.

We measured Prantner's sunspot positions and constructed a butterfly diagram (Figure 5), which shows sunspot groups in both solar hemispheres like Derfflinger's (Hayakawa *et al*., 2020a). Prantner's core sunspot position observations show possible equatorial migration in SC6. Additionally, Prantner's sunspot groups are shown near the maximum of SC5, in the ascending phase of SC7, and immediately after the minimum SCs8/9.

These results strikingly contrast the Dalton Minimum with the Maunder Minimum, in which the few sunspot groups that have been reported (Vaquero *et al*., 2015; Usoskin *et al*., 2015) have positions mostly confined to the southern solar hemisphere (Ribes and Nesme-Ribes, 1993). These observations during the Dalton Minimum appear much more consistent with low-amplitude solar cycles, such as SC24. Prantner's sunspot observations show up to 7–9 sunspot groups and distributions in both solar hemispheres.

Our analyses of Prantner's sunspot observations significantly improve the data for sunspot group number, sunspot positions, and sunspot areas during the Dalton Minimum, and robustly confirm its difference with the Maunder Minimum and similarity with low-amplitude solar cycles in secular minima. Therefore, the Dalton Minimum is not similar to the Maunder Minimum (*e.g*., Kataoka *et al*., 2012; Zolotova and Ponyavin, 2015) but is likely to have a different physical background (*e.g*., Usoskin *et al*., 2015). This conclusion agrees with the significant differences in solar coronal streamers during the Maunder Minimum and the Dalton Minimum (Riley *et al*., 2015; Hayakawa *et al*., 2020b, 2021).

**Acknowledgment**

We thank the Manuscript Department of the Stiftsarchiv Wilten for permitting researches on the MS A07 03 07. We also thank Hannelore Steixner and Dieter Liebmann for their helpful arrangements during HH's visit to the Stiftsarchiv Wilten and fruitful comments/discussions on Stephan Prantner's background. This work was supported in part by JSPS Grant-in-Aids JP15H05812, JP17J06954, JP20K20918, JP20H05643, and JP21K13957, JSPS Overseas Challenge Program for Young Researchers, JSPS Overseas Challenge Program for Young Researchers, the 2020 YLC collaborating research fund and Young Leader Cultivation (YLC) program of Nagoya University, the research grants for Mission Research on Sustainable Humanosphere from the Research Institute for Sustainable Humanosphere (RISH) of Kyoto University, and the Austrian Science Foundation (FWF) project P 31088. We thank SILSO and NASA JPL for managing and providing international sunspot number and JPL DE430. This work was partly merited from participation in the ISWAT-COSPAR S1-01 team and the International Space Science Institute (ISSI, Bern, Switzerland) via the International Team 417 "Recalibration of the Sunspot Number Series".

**Data Availability**

Prantner's original manuscript is preserved as MS A07 03 07 in the Manuscript Department of the Stiftsarchiv Wilten.